\documentstyle[12pt]{article}
\begin{document}

\begin{titlepage}
\rightline{Feb 1998}
\rightline{UM-P-98/13}
\rightline{RCHEP-98/3}
\vskip 2cm
\centerline{\large \bf  
Neutrino masses in the
SU(5)  $\otimes$ SU(5)' mirror symmetric model }
\vskip 1.1cm
\centerline{M. Collie and R. Foot\footnote{Email address:
Foot@physics.unimelb.edu.au}}
\vskip .7cm
\centerline{{\it Research Centre for High Energy Physics}}
\centerline{{\it School of Physics}}
\centerline{{\it University of Melbourne}}
\centerline{{\it Parkville 3052 Australia}}
\vskip 2cm

\centerline{Abstract}
\vskip 1cm
\noindent
Motivated by the atmospheric and solar neutrino
anomalies, we study neutrino masses in a parity 
invariant $SU(5)\otimes SU(5)'$ grand unified model.
Constraints from photon - mirror photon kinetic
mixing are discussed.

\end{titlepage}
\noindent
The possibility that parity is a unbroken symmetry 
of nature seems very appealing. 
In order to achieve unbroken parity symmetry it is necessary 
to contemplate the existence of a set
of mirror particles and forces which are isomorphic to the
ordinary particles and forces\footnote{The general idea
seems to have been first discussed by Lee and Yang\cite{ly}.
See also\cite{P} and references there-in for other 
early references.}. The parity symmetry interchanges 
ordinary and mirror particles. 
However, while the 
number of particles is doubled the number of parameters 
is not significantly increased 
(only two additional parameters in the minimal 
exact parity symmetric model with massless neutrinos)\cite{flv0}.
(For example, each of the 
mirror particles will have the same mass as the corresponding 
ordinary particle since the parity symmetry is unbroken\footnote{
It is possible to build mirror models with the parity symmetry
being spontaneously broken. See Ref.\cite{other2} for
some models of this type.}). 
If the neutrinos have mass and there is mass mixing between ordinary
and mirror neutrinos, then the parity symmetry implies
an interesting constraint on the mixing.
Consider for example the first generation neutrino $\nu_e$ and its
mirror partner $\nu'_e$. Since parity interchanges ordinary
and mirror species it follows that the parity eigenstates are
maximal combinations of weak eigenstates, i.e.
$\nu_{\pm} = {1 \over \sqrt{2}}(\nu_e \pm \nu'_e)$.
It is easy to show\cite{P} that the parity eigenstates are also 
mass eigenstates if the parity symmetry is unbroken. Hence,
provided that the mixing between generations is small, as it is
in the quark sector, then the parity symmetry implies that
each of the known neutrinos $\nu_e, \nu_\mu, \nu_\tau$ will
be approximately maximal mixtures of two mass eigenstates\cite{P}.
Thus the concepts of neutrino mass and unbroken parity
symmetry naturally lead to the prediction 
that each of the known neutrinos will
oscillate maximally with an essentially sterile mirror
partner (which we denote by 
$\nu'_e$, $\nu'_{\mu}$, $\nu'_{\tau}$)\cite{P}.

Maximal $\nu_e \to \nu'_e$ oscillations 
with 
\begin{equation}
3 \times 10^{-10} \stackrel{<}{\sim} |\delta m^2_{ee'}|/eV^2
\stackrel{<}{\sim} 10^{-3},
\label{yy}
\end{equation}
leads to an energy independent
$50\%$ reduction of solar neutrinos thereby
explaining the observed solar neutrino deficit\cite{P}.
(The upper bound in Eq.\ref{yy} is from the recent chooz
experiment\cite{chooz}).

Maximal $\nu_{\mu} \to \nu'_{\mu}$
oscillations can explain the atmospheric neutrino anomaly\cite{P}.
Using the recent preliminary SuperKamiokande data\cite{kearns},
Ref.\cite{fvy} finds that maximal $\nu_{\mu} \to \nu'_{\mu}$
oscillations provide a good fit for the 
range\footnote{
Note that if $|\delta m^2_{ee'}|$ happens to be in the range
$5 \times 10^{-5} \stackrel{<}{\sim} |\delta m^2_{ee'}|/eV^2
\stackrel{<}{\sim} 10^{-3}$
then the atmospheric neutrino experiments will also be
sensitive to $\nu_e \to \nu'_e$ oscillations as well
as the $\nu_{\mu} \to \nu'_{\mu}$ oscillations.
See Ref.\cite{bfv} for details.}
\begin{equation}
1.5 \times 10^{-3} \stackrel{<}{\sim} |\delta m^2_{\mu \mu'}|/eV^2
\stackrel{<}{\sim} 1.5 \times 10^{-2}.
\end{equation}
Furthermore there is a significant range of tau neutrino masses
and mixing parameters where these solutions are {\it not}
in conflict with standard Big Bang Nucleosynthesis (BBN)
when the creation of lepton number asymmetry
due to $\nu_{\tau} \to \nu'_{\mu}$ or $\nu_{\tau}
\to \nu'_e$ oscillations is taken into account\cite{rf}.
Note also that the exact parity symmetric model 
is also compatible with the LSND \cite{lsnd} 
$\nu_\mu \to \nu_e$ oscillation signal\cite{P}.

The above solutions to the atmospheric and solar neutrino problems
will be further tested in the near future by the SNO 
and superKamiokande experiments.
The $\nu_e \to \nu_e'$ oscillations will not affect the ratio
of charged to neutral current solar neutrino induced
events at SNO (which is very different from many other models,
such as the $\nu_e \to \nu_\mu$ MSW solution).
The $\nu_{\mu} \to \nu_{\mu}' $ solution to
the atmospheric neutrino anomaly can be distinguished from
the $\nu_{\mu} \to \nu_{\tau}$ solution
in the near future by studying the properties of the $\pi^0$'s 
produced by neutral current interactions at superKamiokande
\cite{papers,kearns}.
For example, if $\nu_{\mu} \to \nu_{\mu}' $ oscillations
are responsible for the anomaly, then there should be
an up-down asymmetry of the $\pi^0$'s which should be
measurable (if the atmospheric neutrino anomaly is
due to $\nu_\mu \to \nu_\tau$ oscillations then
there should be no such up-down asymmetry for the 
pions). There are also other ways to test 
$\nu_\mu \to \nu'_\mu$ oscillation solution\cite{other}.

Besides neutrino physics and 
gravity\footnote{
The cosmological implications of mirror particles has
been discussed in a number of publications,
see Ref.\cite{cosmo} for details.}
the only other ways in which the mirror particles can interact with
the known particles is through Higgs ($\phi$) - mirror 
Higgs ($\phi'$) interactions\cite{flv0} 
\begin{equation}
\lambda  \phi^{\dagger} \phi \phi'^{\dagger}  \phi'
\label{hig}
\end{equation}
and kinetic mixing of the ordinary and mirror photon\cite{h,gl,cg,flv0}
\begin{equation}
\delta F_{\mu \nu}F^{'\mu \nu}.
\end{equation}
At present there is no significant experimental bound 
on $\lambda$\footnote{
Note that there is a BBN bound $\lambda \stackrel{<}{\sim}
3 \times 10^{-6}\sqrt{M_{higgs}/TeV}$ (see Ref.\cite{cb}) 
which can be derived 
by demanding that the Higgs - mirror Higgs quartic interaction
term, Eq.\ref{hig}, does not bring the mirror 
sector into equilibrium with the ordinary sector 
in the early Universe.  }.
The Higgs mirror Higgs interaction will mainly affect 
Higgs production and decay which can
only be tested if and when the Higgs particle is discovered.
The parameter $\delta$ has the effect of giving the mirror
partners to the charged particles an electric charge
proportional to $\delta$\cite{h}.
The most stringent experimental bound  on $\delta$ comes from
orthopositronium decay\cite{gl}.
The photon - mirror photon kinetic mixing
leads to mass mixing between orthopositronium and mirror
orthopositronium. This means that
maximal oscillations between ordinary orthopositronium
and mirror orthopositronium would occur. Since mirror
orthopositronium decays into undetected mirror photons,
it is possible to deduce a bound on $\delta$ by comparing 
the theoretical and experimental results for 
orthopositronium decay. The bound is\cite{gl}
\begin{equation}
\delta  \stackrel{<}{\sim} 4 \times 10^{-7}.
\label{gl}
\end{equation}
A more stringent bound can be obtained if one assumes that
the standard assumptions of cosmology are correct
and this leads to a bound of about
$\delta \stackrel{<}{\sim} 3 \times 10^{-8}$\cite{cg} 
(otherwise the the photon - mirror photon kinetic mixing
will populate the mirror particles and adversely affect BBN).
In the minimal parity symmetric model based on 
the gauge group,
\begin{equation}
SU(3) \otimes SU(2) \otimes U(1) \otimes
SU(3)' \otimes SU(2)' \otimes U(1)'
\label{dfj}
\end{equation}
$\delta$ is a free parameter. Its necessary smallness gives 
a motivation for
extending this theory to the grand unified version, e.g.
$SU(5) \otimes SU(5)'$\cite{gl}. In this case a tree level kinetic
mixing term is forbidden by the gauge symmetry.
(Gauge kinetic mixing is only gauge invariant for abelian
$U(1)\otimes U(1)'$ symmetry).

In the usual SU(5) model\cite{gg}, denote the fermion $\bar 5$ 
and $10$ representations as
\begin{equation}
f_{1L} \sim \bar 5, \   f_{2L} \sim 10
\end{equation}
and the Higgs scalar responsible for electroweak
symmetry breaking as 
\begin{equation}
\phi \sim \bar 5.
\end{equation}
In the parity symmetric 
$SU(5)\otimes SU(5)'$ model, in addition to the ordinary
fermions and Higgs scalars (which are $SU(5)'$ singlets), 
there are mirror fermions and mirror
scalars (which are $SU(5)$ singlets) and transform under 
$SU(5)'$ as:
\begin{equation}
f'_{1R} \sim \bar 5, \  f'_{2R} \sim 10, \ 
\phi' \sim \bar 5.
\end{equation}
The parity transformation interchanges the ordinary particles
with their mirror counterparts (i.e.
$f_{1L} \leftrightarrow \gamma_0 f'_{1R}, \
f_{2L} \leftrightarrow \gamma_0 f'_{2R}, \
\phi \leftrightarrow \phi'$ and similarly for the gauge bosons).
Note that we assume that $\langle \phi \rangle = \langle \phi'
\rangle$ so that parity is unbroken by the vacuum.
(It is straight
forward to show that this happens for a range of parameters of the
Higgs potential, without the need for additional scalar particles.
We also assume that the GUT scale breaking of 
$SU(5)\otimes SU(5)'$ down to Eq.\ref{dfj} respects 
the parity symmetry.)

In order to incorporate neutrino masses we need to introduce either
gauge singlet neutrinos or introduce
additional scalar particles. The addition of one gauge singlet
neutrino per generation (and the corresponding mirror neutrinos)
leads essentially to the usual see-saw model\cite{ss}.
Denote the gauge singlet neutrinos as
$\nu_{R}$ and the mirror singlet as $\nu'_{L}$
(here and elsewhere the generation index is suppressed).
Note that $\nu_{R} \leftrightarrow \gamma_0 \nu'_{L}$ under 
the parity transformation.
The Yukawa Lagrangian consists of the usual terms giving
masses to the charged fermions (and
the corresponding terms for the mirror charged fermions) 
together with the terms:
\begin{equation}
{\cal L}_{yuk}^{\nu} = \lambda_1 \bar f_{1L} \phi \nu_{R} +
\lambda_1 \bar f'_{1R} \phi' \nu'_{L} 
+ \lambda_2 \bar f_{1L}\phi (\nu'_{L})^c +
\lambda_2 \bar f'_{1R}\phi' (\nu_{R})^c + H.c..
\end{equation}
The above Yukawa Lagrangian gives Dirac mass terms
to the neutrinos after spontaneous symmetry breaking.
If the gauge singlet neutrinos have
large bare masses:
\begin{equation}
{\cal L}_{bare} = 
M_1\bar \nu_{R}(\nu_{R})^c + M_1\bar \nu'_{L}(\nu'_L)^c
+ M_2(\bar \nu_{R}\nu'_{L} + \bar \nu'_L \nu_R) + H.c.,
\end{equation}
then provided that $M_1, M_2 \gg \langle \phi \rangle 
= \langle \phi' \rangle \equiv u \simeq 170\ GeV$, the 
neutrinos develop small Majorana masses (which
is just the usual see-saw mechanism).
Furthermore, if the mixing between the generations is small
then each of the weak eigenstates are an approximately
maximal mixture of two mass eigenstates (which is an automatic
consequence of the parity symmetry).

Alternatively, neutrino masses can be incorporated without
adding gauge singlet fermions by extending
the scalar sector.
In this case we need to add the following multiplets
\begin{equation}
\rho \sim (15, 1), \ \rho' \sim (1, 15), \
\chi \sim (\bar 5, 5),
\end{equation}
with $\rho \leftrightarrow \rho'$ and $\chi \leftrightarrow
\chi^{\dagger}$ under the parity symmetry.
The scalars couple to the fermions with
\begin{equation}
{\cal L}_{yuk} =
\lambda_1 \bar f_{1L} \rho^{\dagger} (f_{1L})^c +
\lambda_1 \bar f'_{1R} \rho'^{\dagger}(f'_{1R})^c +
\lambda_2 \bar f_{1L} \chi f'_{1R} + H.c.
\label{hap}
\end{equation}
The most general Higgs potential contains the terms:
\begin{equation}
V_1 = m_{\chi}^2 \chi^{\dagger}\chi + 
m_{\rho}^2(\rho^{\dagger} \rho + \rho'^{\dagger} \rho')
+ m_1 (\phi^2 \rho + \phi'^2 \rho')
 + m_2 (\phi^{\dagger} \phi' \chi) + H.c.
\end{equation}
If the trilinear terms are neglected then the 
vacuum expectation values (VEVs) of $\chi, 
\rho$ are trivially zero (we are assuming that
$m_{\chi}^2, m_{\rho}^2 > 0$). However given that 
$\langle \phi \rangle = \langle \phi' \rangle \equiv u$,
the effects of the trilinear term is to induce linear terms
in $\rho, \rho', \chi$ which destablises the vacuum (this
result is quite general, and will hold for the most general
Higgs potential for a range of parameters).
Because of this, small VEVs for $\chi$ and $\rho, \rho'$ arise
\begin{equation}
\langle \rho \rangle = \langle \rho' \rangle = m_1 u^2/m_{\rho}^2, \
\langle \chi \rangle = m_2 u^2/m_{\chi}^2.
\end{equation}
These VEVs are naturally much smaller than the electroweak
symmetry breaking VEV ($\langle \phi\rangle = \langle
\phi' \rangle = u$) 
provided that $m_\chi, m_\rho \gg u$.
This is achieved without any new hierarchy problem.
(There is still
the old hierarchy problem, i.e. why $\langle \phi \rangle$ 
is so small c.f. GUT scale).
The smallness of the VEV is related to the high mass scale of the
$\rho, \chi$.  For example, if $m_1 \sim m_{\rho} 
\approx 10^{10} \ GeV$
then $\langle \rho \rangle/\langle \phi \rangle \approx 10^{-8}$.
Clearly, these VEVs are naturally small and thus
so are the induced Majorana neutrino masses (from Eq.\ref{hap}).
As expected from the parity symmetry, the usual weak
eigenstate neutrinos are each maximal combinations of
two mass eigenstates (if mixing between the generations is
neglected).
Note that without the $\chi$ scalar, there is no mass mixing 
between the ordinary and mirror neutrinos so no ordinary
neutrino - mirror neutrino oscillations would
occur. Since we are interested in using the
maximal neutrino - mirror neutrino oscillations to
explain the solar and atmospheric neutrino puzzles, we 
must include the $\chi$ scalar.

Notice that the $\chi \sim (\bar 5,5)$ 
couples to both ordinary and mirror
photons so that kinetic mixing of the ordinary and mirror
photons will be radiatively induced.
Such radiatively induced kinetic mixing was first studied by
Holdom\cite{h} in quite a general context and applied to the
$SU(5) \otimes SU(5)'$ model by Glashow\cite{gl}.
Under the low energy subgroup, Eq.\ref{dfj},
$\chi \to \chi_1, \chi_2, \chi_3, \chi_4$ where
\begin{eqnarray}
\chi_1 &\sim (3,1,1/3,3,1,1/3),\ 
\chi_2 \sim (3,1,1/3,1,2,-1/2), \nonumber \\
\chi_3 &\sim (1,2,-1/2,3,1,1/3), \
\chi_4 \sim (1,2,-1/2,1,2,-1/2).
\end{eqnarray}
If these multiplets have
masses $m_i, i=1,2,3,4$ (note that $m_2 = m_3$
due to the parity symmetry) then the
1-loop radiative contribution to the photon - mirror photon kinetic 
mixing is\cite{h,gl}
\begin{equation}
\delta = {\alpha \over 2\pi}log {m_1 m_4 \over m_2^2}.
\end{equation}
Notice that this scenario is in conflict with Glashow's bound
Eq.\ref{gl} unless $m_1 m_4 \simeq m_2^2$.
Thus it seems unlikely that particles such as $\chi$
which couple to both ordinary and mirror particles could
exist (this was essentially the conclusion of Ref.\cite{gl,cg}).
Nevertheless, we suggest that such
particles cannot be definitely excluded. 
One possibility is that
$\chi_i$ are approximately degenerate at tree level
(to within $10^{-4}$) which occurs for a range
of parameters in the most general Higgs potential.
For example this will happen if the $\chi$ couples weakly to the
other scalar particles. 
If the $\chi_i$ are degenerate at tree level then the one - loop 
contribution to $\delta$ vanishes and the leading
contribution to $\delta$ is naively
expected to arise at  2-loops in perturbation theory.
The only radiative corrections which do not depend on some
arbitrary Yukawa couplings are the
radiative corrections involving virtual  
$SU(5)\otimes SU(5)'$ gauge bosons.
Observe that
the leading non-zero contribution to $\delta$ involving virtual
$SU(5)\otimes SU(5)'$ gauge bosons 
 must contain both ordinary and mirror virtual 
gauge bosons.
To see this, observe that if say mirror 
gauge bosons are neglected and the $\chi_i$ are degenerate 
at the tree-level then the interaction Hamiltonian 
describing the interactions of the $\chi$ with the 
ordinary gauge bosons is invariant under mirror $SU(5)'$
symmetry. This symmetry will forbid the photon - 
mirror photon kinetic mixing.
Thus the leading gauge contribution to $\delta$
{\it must} involve virtual $SU(5)$ gauge bosons {\it and} virtual
$SU(5)'$ gauge bosons.
It follows that the leading gauge contribution to $\delta$
must be of order $(\alpha/\pi)^3$ or 
higher\footnote{
This result holds quite generally. For example imagine that
there exists heavy exotic fermions of the form:
$F_L \sim (5,5),\ F_R \sim (5,5)$. Then the mass term
${\cal L}_{mass} = M\bar F_L F_R + H.c.$ would give
all the components of $F$ degenerate masses at tree-level.
Such fermions could only contribute to the photon - mirror
photon kinetic mixing at the three loop level (or higher).
Similar results would also occur in other GUT models, such
as $SO(10)\otimes S0(10)$.}.
Thus the leading contribution to $\delta $ is
\begin{equation}
\delta \stackrel{<}{\sim} {\cal{O}}\left[ (\alpha/\pi)^3 \right]
\sim 10^{-6} - 10^{-8}.
\end{equation}
Thus, the quite stringent bound on the photon - mirror photon 
kinetic mixing does not exclude the possible 
existence of  $\chi \sim (\bar 5,5)$ 
scalars provided that they are 
degenerate at tree level (which will be a good approximation 
for a range of parameters).
Furthermore it seems likely that the kinetic mixing, if
it arises at three-loops could be close to the experimental limit 
(i.e. within an order of magnitude or two)
and thus should be testable in future experiments.

We conclude by summarising the ideology of this paper.
The success of the exact parity symmetry in explaining the
observed neutrino physics anomalies together with
the rather stringent bound on the photon - mirror
photon kinetic mixing motivates consideration of grand unified
mirror symmetric GUT models with neutrino masses.
We have considered the parity symmetric $SU(5) \otimes SU(5)'$
model for definiteness. Two distinct ways to incorporate 
neutrino masses in this model are envisaged.  One 
way involves adding a 
gauge singlet fermion to each generation. In this 
case the usual see-saw neutrino masses
results and the model reduces to the exact parity
symmetric model (with small neutrino masses) at low 
energies (with zero kinetic
mixing due to the underlying GUT symmetry of the
Lagrangian).  The other way of incorporating neutrino 
masses is to extend the scalar sector.
This possibility suggests that the
photon - mirror photon kinetic mixing is
non-zero since it is generated radiatively.
We have argued that the kinetic mixing in such models
may well be close to the experimental limit.
\vskip 0.8cm
\noindent
{\bf Acknowledgement}
\vskip 0.4cm
\noindent
We would like to thank Ray Volkas for discussions.
R.F. is an Australian Research Fellow.

\end{document}